\documentclass{PoS}

\usepackage{url}

\def\arxiv#1{\href{http://arxiv.org/abs/#1}{{\tt arXiv:#1}}}

\def\link#1{\href{#1}{\tt#1}}

\title{Multimessenger studies with the VERITAS Atmospheric Cherenkov Observatory}

\ShortTitle{Multimessenger studies with VERITAS}

\author{\speaker{Marcos Santander}, for the VERITAS Collaboration\thanks{{http://veritas.sao.arizona.edu/}}\\
        Barnard College, Columbia University \\
        E-mail: \email{santander@nevis.columbia.edu}}

\abstract{The VERITAS gamma-ray observatory has an active multimessenger program, currently focused on studying the connection between very-high-energy gamma rays and the astrophysical neutrino flux recently discovered by the IceCube telescope. As both gammarays and neutrinos are produced in hadronic interactions, it is expected that a joint study of both messenger channels could reveal powerful cosmic-ray accelerators and probe their properties. We present an overview and recent results from the VERITAS multimessenger program and discuss prospects for combined studies with other multimessenger facilities, such as the Advanced LIGO gravitational wave observatory.}

\FullConference{38th International Conference on High Energy Physics\\
		3-10 August 2016\\
		Chicago, USA}

\begin{document}

\section{Introduction}

The discovery of astrophysical neutrinos by IceCube and the recent announcement of the detection of gravitational waves by the LIGO collaboration open new possibilities for the study of the universe by combining these observations with those from existing electromagnetic instruments. Multimessenger astronomy can address several important questions in high-energy astrophysics such as the origin of cosmic rays, the nature of high-energy emission in transient astrophysical objects, and the localization and characterization of compact object mergers, among others.

The VERITAS multimessenger program is currently focused on performing follow-up observations of astrophysical neutrino candidate events reported by IceCube to search for transient and steady sources of gamma rays that could be associated with the neutrino emission. We describe here the status of the neutrino follow-up program of VERITAS and present some preliminary results from this search. We also present prospects for follow-up observations of gravitational wave alerts from the LIGO and Virgo observatories.

VERITAS~\cite{VTS} is a ground-based instrument for gamma-ray astronomy with maximum sensitivity in the 80 GeV to 30 TeV range. It consists of an array of four 12-m optical telescopes each equipped with a camera containing 499 photomultiplier tubes (PMTs) covering a field of view of $3.5^{\circ}$. The array is located at the Fred Lawrence Whipple Observatory (FLWO) in southern Arizona (31$^{\circ}$ 40'N, 110$^{\circ}$ 57'W, 1.3km a.s.l.). The angular resolution of VERITAS is $\sim0.1^{\circ}$ at 1 TeV (for 68\% containment) and the energy resolution is $\sim20$\% at the same energy.

\section{IceCube Neutrino Follow-up Observations}

The discovery of an astrophysical flux of high-energy neutrinos by IceCube \cite{HESE1} is a major step 
towards finding the origin of cosmic rays, as neutrinos are produced in hadronic interactions taking place in cosmic-ray accelerators. The astrophysical neutrino flux is significant at energies between $\sim20$ TeV and a few PeV and its spectrum is consistent with an $E^{-\Gamma}$ power-law, with different IceCube analyses and neutrino detection channels showing $\Gamma$ to be in the 2.1--2.5 range~\cite{Leif}. So far, no neutrino point-sources or significant correlation with the Galactic plane have been identified that could reveal the origin of the emission, although the apparent isotropy of the neutrino events collected thus far seems to favor a dominant extragalactic component. 

The hadronic interactions that give rise to the neutrino emission should also lead to the production of gamma rays through neutral pion decays. Unless the sources or the propagation medium are optically thick to high-energy gamma rays, this flux could be detected at Earth using instruments such the VERITAS telescope array. Combining neutrino and gamma-ray observations increases the sensitivity of the search for hadronic sources. The sensitivity of  this correlation study depends on the angular resolution of the neutrino directional reconstruction. In IceCube, only events associated with $\nu_{\mu}$ charged-current interactions are reconstructed with a typical angular uncertainty of $1^{\circ}$ or better due to the km-long muon tracks they produce in the detector. 

Over the last years, VERITAS has been used to search for gamma-ray emission associated with IceCube high-energy muon neutrino events, as their $\sim1^{\circ}$ error region can be covered with the VERITAS field of view. The location of VERITAS in southern Arizona restricts the observable neutrino positions to those in the Northern sky, or at low declinations in the Southern sky. The VERITAS follow-up program originally focused on muon positions released by IceCube in publications or through private communications, which introduced a delay of several months between the neutrino detection and gamma-ray observations. We describe preliminary results from these observations in Subsection~\ref{subsec:archival}. In April 2016, IceCube began circulating public alerts for interesting neutrino events minutes after they are detected at the South Pole as a way of improving the sensitivity of electromagnetic follow-ups to transient sources. We present preliminary results from a VERITAS prompt follow-up of a ``realtime" alert in Subsection~\ref{subsec:realtime}.

\subsection{Observation of archival neutrino positions}\label{subsec:archival}

\begin{figure}[h]
\centering
\includegraphics[width=0.65\textwidth]{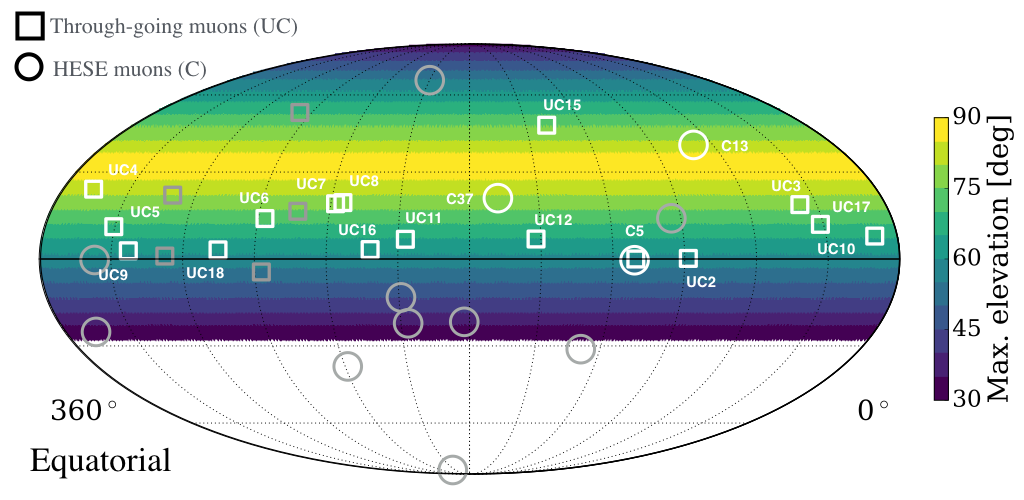}
\caption{Skymap in equatorial coordinates showing the positions of HESE (`C', circles) and through-going (`UC', squares) muon events compared to the maximum elevation these sources reach when observed from VERITAS. 
Typical observations are performed when sources go above $50^{\circ}$ elevation. White markers indicate those positions where VERITAS has collected observations.}
\label{vts_map}
\end{figure}

Muon neutrino positions observable with VERITAS were selected from three IceCube data sets: six muon events from a sample of high-energy starting events (HESE) collected by IceCube over four years~\cite{HESE3},  21 events from a sample of through-going Northern-sky muon tracks collected over two years~\cite{Chris}, and 29 through-going tracks with high-astrophysical purity collected over six years~\cite{Leif}, with some overlap between the different data sets. The map in Fig.~\ref{vts_map} shows the position of the muon events selected for observation.

Thus far, VERITAS has collected a total of 57 hours of good-quality data on 18 muon neutrino positions. The observations have been performed using the standard \emph{wobble} observation strategy where the telescopes are offset from the position of the potential source to allow for a simultaneous determination of the background. Offsets of $0.5^{\circ}$ and $0.7^{\circ}$ with respect to the best fit neutrino location were used to provide better coverage of the neutrino error circle. During the analysis of VERITAS data, cuts are introduced to separate gamma-ray shower candidate events from a dominant background of hadronic cosmic-ray showers. In this work, we have used \emph{soft} cuts optimized for sources with spectral indices of $\sim -4$. 
No significant gamma-ray excess has been found in these observations and  consequently 99\% confidence-level upper limits have been derived at the neutrino positions. Preliminary integral upper limits above 100 GeV are given in Table~\ref{tab:uls} for a subset of the observed neutrino positions. On average, the limits are at the level of a few percent of the Crab nebula gamma-ray flux.

\begin{table*}[h]
\small
\centering
\begin{tabular}{c | c | c | c}
ID  & Observation time [min] & UL (99\%) [cm$^{-2}$ s$^{-1}$] & UL (99\%) [C.U.]\\
\hline
C5 & 180 & $8.33 \times 10^{-12}$ & 2.3\%\\
C13 & 574 & $4.01 \times 10^{-12}$ & 1.1\% \\
C37 & 275 & $7.30 \times 10^{-12}$ & 2.0\% \\
\hline
\hline
UC2 & 25 & $2.12 \times 10^{-11}$ & 5.8\% \\
UC3 & 180 & $6.31 \times 10^{-12}$ & 1.7\% \\
UC4 & 122 & $9.89 \times 10^{-12}$ & 2.7\% \\
UC5 & 90 &  $6.66 \times 10^{-12}$ & 1.8\% \\
UC6 & 25 & $9.53 \times 10^{-12}$ & 2.6\% \\
UC7 & 15 & $3.96 \times 10^{-11}$ & 10.9\% \\
UC8 & 60 & $9.31 \times 10^{-12}$ & 2.6\%\\
UC9 & 40 & $1.52 \times 10^{-11}$ & 4.2\%\\
UC10 & 90 & $9.40 \times 10^{-12}$ & 2.6\% \\
UC11 & 209 & $4.4 \times 10^{-12}$ & 1.2\% \\
UC12 & 25 & $9.53 \times 10^{-12}$ & 2.6\% \\
UC15 & 90 & $7.40 \times 10^{-12}$ & 2.0\% \\
UC16 & 40 & $8.57 \times 10^{-12}$ & 2.4\% \\
UC17 & 150 & $4.41 \times 10^{-12}$ & 1.2\%\\
UC19 & 210 & $3.92 \times 10^{-12}$ & 1.1\% \\
\end{tabular}
\caption{Summary of the observations of high-energy muons. The table follows the notation given in~\cite{ICRC2015}: HESE event IDs start with `C' and through-going tracks with `UC'. Upper limits at 99\% level are given in units of integrated flux and as a percentage of the Crab nebula flux above the same energy threshold of 100 GeV.}
\label{tab:uls}
\end{table*}

\subsection{Prompt follow-up observations of neutrino event alerts}\label{subsec:realtime}

On April 8th, 2016 IceCube started broadcasting public alerts for high-energy HESE muon events. On average, $\sim3.5$ neutrino alerts are expected per year, with one being astrophysical in origin.  The angular resolution for these events is in the 2$^{\circ}$ to 8.9$^{\circ}$ range. Alerts are issued by the Astrophysical Multimessenger Observatory Network (AMON)\footnote{\link{http://amon.gravity.psu.edu/}} and circulated using the Gamma-ray Coordinates Network (GCN)\footnote{\link{http://gcn.gsfc.nasa.gov/}}. The typical alert latency, between the neutrino detection and the submission of the alert message, is below three minutes. VERITAS is subscribed to the HESE alert stream and receives these notices in electronic form which can be parsed and interpreted by the data acquisition software. Upon reception, observers are notified and telescopes are repointed to the neutrino location to start follow-up observations.

\begin{figure}[h]
\centering
\includegraphics[width=0.65\textwidth]{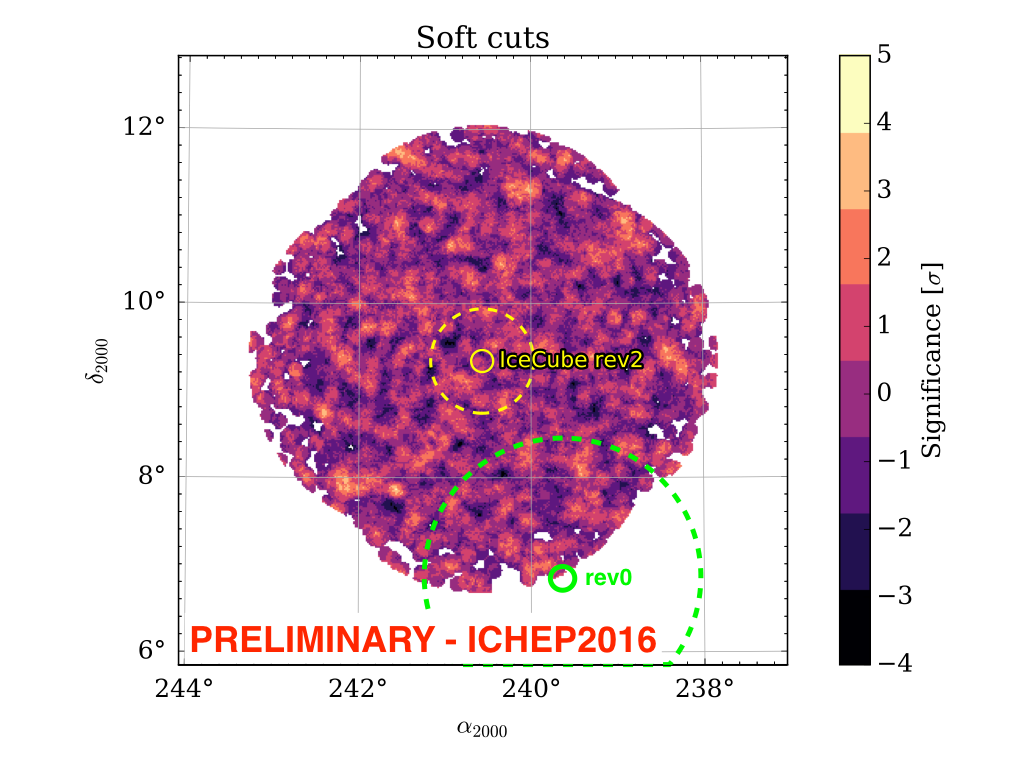}
\caption{Significance skymap for observations taken on the revised neutrino position for the IceCube realtime muon alert. The first position circulated by IceCube and its uncertainty are shown in green.}
\label{fig:realtime}
\end{figure}

The first alert from this system was sent out on April 27th, 2016 at 05:53:53 UTC when a neutrino, detected at 05:52:32 UTC, was reported from the coordinates RA=239.66$^{\circ}$, Dec=+6.85$^{\circ}$~\cite{AMON1}. VERITAS was operating at the time and started observations at 05:55:45 UTC, less than four minutes after the neutrino detection. A total of 71 minutes of good-quality data were taken on the alert position under bright moonlight during the first night. Based on an offline analysis performed by IceCube the following day, the neutrino position was revised to RA=240.57$^{\circ}$, Dec=+9.34$^{\circ}$, outside of the area covered by the observations taken during the first night. During the second night, 118 minutes of good-quality data were taken in bright moonlight conditions on the revised location. No gamma-ray excess was identified in VERITAS observations and upper limits were circulated to the community in a GCN circular~\cite{GCNAMON1}. A significance sky map for the observations taken on the revised position is shown in Fig.~\ref{fig:realtime}.

In July 2016 IceCube added a second alert stream for extremely-high energy (EHE) muon events with a higher astrophysical probability (on average, two out of the four sent every year will be expected to be astrophysical) and a better angular resolution in the 0.1$^{\circ}$-$0.4^{\circ}$ range. The first alert for this system was sent out on July 31st, while VERITAS was not operational.

\section{Prospects for gravitational wave follow-up observations}

\begin{figure}[h]
\centering
\includegraphics[width=0.7\textwidth]{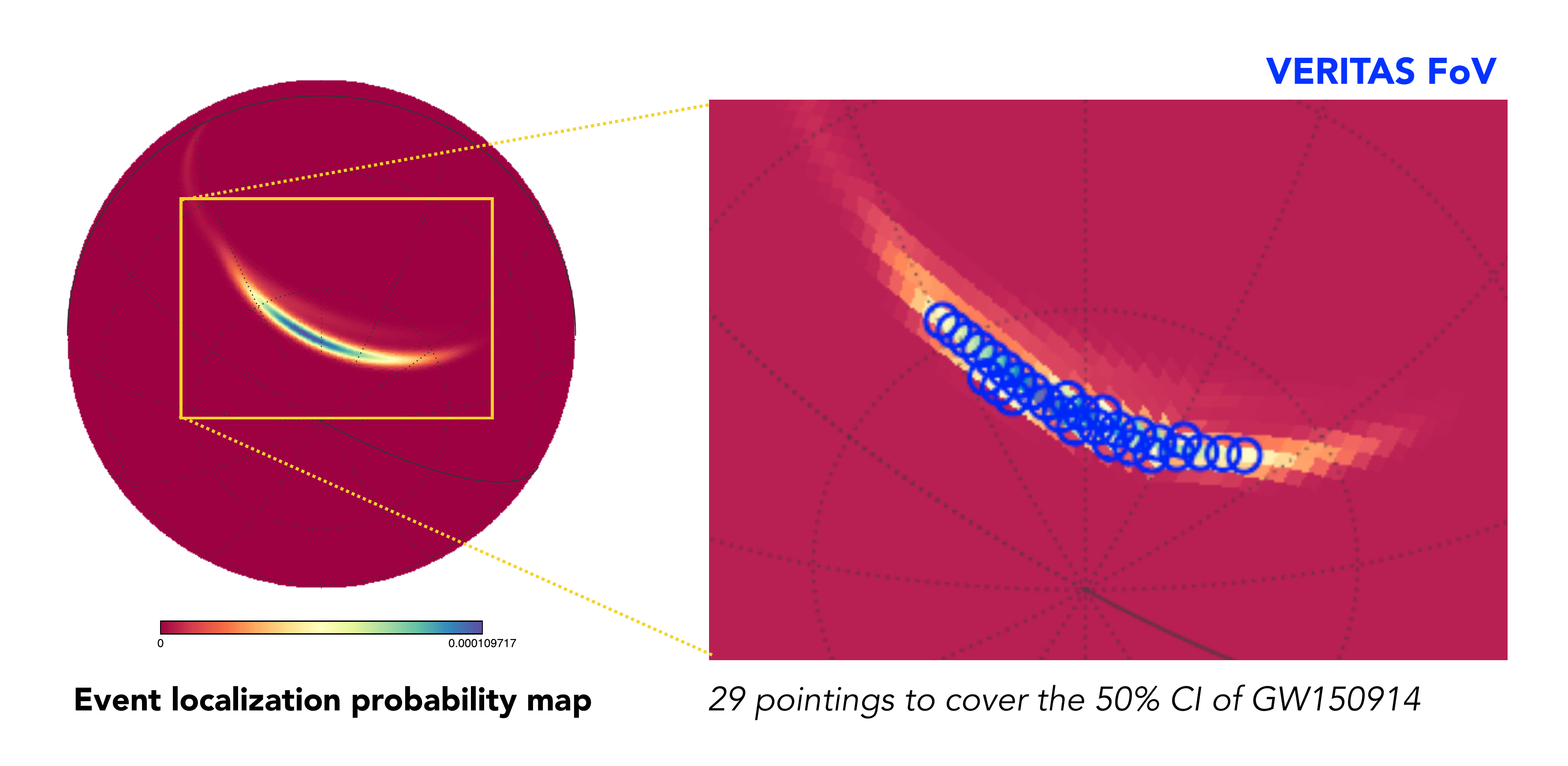}
\caption{Proposed strategy for the VERITAS tiling of a future GW alert. We use the GW discovery event,  GW150914, as an example to show that the 50\% containment region for the alert could be covered in 29 pointings.}
\label{fig:gw}
\end{figure}

The recent discovery of gravitational waves by the LIGO collaboration~\cite{LIGO} brings new opportunities for multimessenger studies. Compact object mergers that lead to gravitational wave (GW) emission may also be responsible for gamma-ray bursts and therefore a gamma-ray follow-up could reveal the location of the GW source. The main challenge for VERITAS follow-up observations of GW alerts is the large uncertainty regions associated with the events, of $\mathcal{O}$(100 deg$^{2}$). The $\sim10$ deg$^{2}$ VERITAS field of view can be used to tile the alert uncertainty region and set constraints on the gamma-ray emission associated with a GW event. A possible tiling strategy is shown in Fig.~\ref{fig:gw}.

Prospects for GW localizations will improve in 2017 thanks to the restart of Advanced LIGO and the beginning of  Advanced Virgo observations. The two instruments combined could reduce the size of the containment regions to  $\mathcal{O}$(10 deg$^{2}$), facilitating VERITAS observations.

\section{Conclusions and future outlook}

VERITAS operates an active multimessenger program currently focused on performing follow-up observations of IceCube neutrino events. Electromagnetic follow-up observations of neutrino events continues to be one of the most promising way of finding the sources of the neutrino flux detected by IceCube. No gamma-ray counterpart has been identified so far in VERITAS observations of 18 high-energy neutrino positions. The effort is now shifting to concentrate on realtime alert follow-ups which increase the sensitivity of this search to transient sources. This program will expand in the near future to perform follow-up observations of gravitational wave events. 

\section[]{Acknowledgements}
VERITAS research is supported by grants from the U.S. Department of Energy Office of Science, the U.S. National Science Foundation and the Smithsonian Institution, and by NSERC in Canada. We acknowledge the excellent work of the technical support staff at the Fred Lawrence Whipple Observatory and at the collaborating institutions in the construction and operation of the instrument.
The VERITAS Collaboration is grateful to Trevor Weekes for his seminal contributions and leadership in the field of VHE gamma-ray astrophysics, which made this study possible.


\begin{thebibliography}{99}
\small
\bibitem{VTS}
J.~Holder et al. (VERITAS Collab.), Astropart. Phys. 25, 391-401 (2006)

\bibitem{HESE1} 
M.~G. Aartsen et al. (IceCube Collab.), 
Science 22 (2013), 342, 6161 \arxiv{1311.5238}.


\bibitem{Leif}
M.~G. Aartsen et al. (IceCube Collab.),
Submitted to ApJ, \arxiv{1607.08006}.

\bibitem{HESE3} 
M.~G. Aartsen et al. (IceCube Collab.),
Proc. of the ICRC (2015) \arxiv{1510.05223}.

\bibitem{Chris}
M.~G. Aartsen et al. (IceCube Collab.), 
PRL 115, 081102 (2015) \arxiv{1507.04005}.

\bibitem{ICRC2015} 
M.~Santander (for the VERITAS and IceCube Collab.),
Proc. of the ICRC (2015) \arxiv{1509.00517}.

\bibitem{AMON1}
\link{http://gcn.gsfc.nasa.gov/notices\_amon/67093193\_127853.amon}

\bibitem{GCNAMON1}
\link{http://gcn.gsfc.nasa.gov/gcn3/19377.gcn3}

\bibitem{LIGO}
B. P. Abbott et al. (LIGO and Virgo Collaborations), PRL 116, 061102 (2016)




\end{thebibliography}
\end{document}